# Uncertainty Assessment of Probabilistic Cellular Automata Simulations in Microstructure Evolution

Majid Seyed-Salehi [1*]

1. Faculty of Materials Science and Engineering, K. N. Toosi University of Technology, 1969764499, Tehran, Iran.

    * Corresponding author: Majid Seyed-Salehi, e-mail: seyedsalehi@kntu.ac.ir.

**Abstract**

The probabilistic cellular automaton (PCA) method is highlighted for its relatively simple numerical algorithm and low computational cost in the simulation of microstructural evolution. In this method, probabilistic state change rules are implemented to compute the evolution of cell states at each time step. The stochastic nature of this simulation method leads to non-repeatable simulation results, introducing inherent uncertainty. In this study, the uncertainty and dispersion in PCA simulations of microstructural evolution were investigated. Hence, the probabilistic transformations of cell states were meticulously considered at each time step, and discrete probability distribution functions (dPDF) were introduced to analyze the frequency distribution of simulation outcomes. To evaluate the performance of the proposed dPDFs, cellular automaton models were developed with various numbers of cells and distribution of transformation probabilities. Multiple iterations of these simulations were conducted, and the validity of the presented distribution functions was assessed through statistical analysis of the simulations' outcomes. Comparisons between PCA simulation results and distribution functions demonstrate consistency, emphasizing the predictive capability of the proposed models. Also, the effects of modeling parameters on the uncertainty of simulation results in two and three-dimensional PCA modeling were studied, introducing the coefficient of variation as a measure of dispersion. Results indicate that increasing the number of boundary cells, cellular resolution, and model size reduces uncertainty, enhancing the repeatability of PCA simulation outcomes.

**Keywords**:

Probabilistic Cellular Automata, Probability Distribution Function, Uncertainty Assessment, Microstructure Evolution.

# 1. Introduction

Various numerical methods, including Monte Carlo, cellular automaton (CA), phase field, level-set, and vertex method, have been employed to simulate microstructural evolution. Cellular automation has gained significant interest due to its low computational cost, relatively simple numerical algorithm, and satisfactory computational accuracy [1, 2]. In cellular automaton, the states of individual system components (cells) are determined in discrete spatial and temporal domains using either deterministic or probabilistic rules. In this method, complex physical systems can be explained by defining relatively simple concepts and rules to describe the state transformation of cells and their interconnections [3]. Using probabilistic rules in the CA method simplifies the numerical algorithm and reduces the computational cost. For instance, In the simulation of microstructural evolutions controlled by interface movement, the transformation probability of each boundary cell is computed using governing kinetic equations of boundary motion. Subsequently, a decision is made regarding cell transformation, which can only exhibit two possible states: success or failure. In other words, the boundary either remains stationary or moves to a specific extent. Nevertheless, in reality, the interfaces move continually during transformation.

In material science, the PCA is used in simulating various phenomena, such as solidification, Grain growth, and static and dynamic recrystallization. This method has been widely used for simulation of microstructural evolution during recrystallization. In the late 1990s, Raabe [4, 5] modeled the probabilistic nature of the grain boundary migration activated by atomic thermal fluctuations in the recrystallization phenomenon. Subsequently, the attack frequency (switching frequency of cell) and the probability of transformation of each boundary cell are derived by employing principles of cellular automaton modeling. He introduced a scalable kinetic probabilistic cellular automaton method, utilizing a weighted random sampling Monte Carlo to simulate the stochastic nature of the boundary dynamics during recrystallization. The PCA method has been used to predict both nucleation and growth of new grains in static [1, 6-8] and dynamic [9-11] recrystallization. Moreover, this method is extensively employed to simulate the grain growth phenomenon, considering the structure, curvature, energy, and misorientation of grain boundaries. In these researches, various aspects of microstructure, including grain growth kinetics, grain size distribution, grain morphology, and abnormal grain growth, have been investigated using 2D [12, 13] and 3D [14, 15] models. Rappaz et al. [16] employed the PCA method to simulate the dendritic grain structures in solidification. Recently, this method has been used in the simulation of various aspects of solidification, including epitaxial growth, competitive growth, and dendritic growth in additive manufacturing [17-20].

In the PCA method, the transformation probability is computed as the ratio of the local boundary velocity to the maximum boundary velocity in the domain (or the ratio of the local boundary movement to the maximum admissible boundary movement). Thus, the probabilistic state change rule, determining the new state of the cell *i* in the *k+1* time step ($\xi_i^{k+1}$), is defined based on the state of this cell and its neighbors in the $k^{th}$ time step as below [2].

$$\xi_i^{k+1} = \begin{cases} \xi_i^k & if\ r > \omega_i \quad (Transformation\ Refused) \\ New\ state & if\ r \leq \omega_i \quad (Transformation\ Succeed) \end{cases} \quad (1)$$

$$\omega_i = \frac{v_i}{v_{max}} \quad v_{max} = \max_{\forall j \in \Omega}(v_j)$$

where $\omega_i$ and $v_i$ are the switching probability and boundary velocity of cell $i$, and $v_{max}$ is the maximum switching probability in domain Ω. Here $r$ is a random number ($r \in \mathbb{R}\ |\ 0 \leq r \leq 1$). This rule has been applied, either in the same manner or with minor modifications in numerous studies employing probabilistic cellular automaton [5, 21].

Generally, numerical simulation involves four main steps i.e., developing a conceptual model and proposing mathematical governing equations, developing numerical algorithms to solve the governing equations, implementing the computer program, and finally, validation and verification of model and simulation [22]. In the final step, model validation and verification are performed to assess the accuracy and reliability of the numerical simulation. Model verification evaluates the correctness of the model's assumptions, implementation, and technical specifications by comparing the simulation results with theoretical or analytical results. Validation assesses the model's capability to reproduce real-world observations or experimental data with adequate accuracy [23]. In most microstructural simulation studies, the typical procedure involves introducing the model, conducting the numerical simulations, and subsequently comparing the obtained results with experimental data to validate the model. However, in these researches, it is assumed that the mathematical model and numerical algorithms are already confirmed, and no uncertainty is attributed to models and algorithms. So, in most microstructural simulation studies, the focus lies on assessing the accuracy of the simulation by comparing the obtained results with experimental observations.

Similar to other simulation methods, the solution uncertainties and simulation verification in CA simulations have not received significant attention despite the extensive research conducted in the field. Only a few studies have focused briefly on the uncertainties associated with CA [24-26]. Chen et al. [24] conducted the CA to simulate the prey-predator ecological systems and studied the effects of cell size and cellular configurations in the dynamics of systems. The findings revealed that cell size has significant effects on spatial patterns, while cellular configurations affect both spatial patterns and system stability. Sitko et al. [25] studied the reliability in the predictions of the CA static recrystallization model at different levels of CA space discretization and time step length. They proposed threshold values for CA cell size and time step length to obtain reliable predictions. The findings showed a direct correlation between the accuracy of the simulation results and the resolution of the spatial and temporal domains. Yeh et al. [26] employed CA to simulate the evolution of urban systems. They investigated the effects of GIS (geographic information system) data errors and modeling uncertainties on urban simulation. It showed that the CA model generated relatively stable simulation results at the macro-level despite variations observed at the micro-level. It is also recommended to calculate the probability maps by repeating the numerical simulation and overlaying the repeated simulation results. The researchers studied the effects of various factors, such as cell size and structure resolution [24, 25, 27], time step length [25], cellular configurations and neighborhood definition [24, 28], state change law, initial conditions [26, 29-32] and stochastic nature of PCA algorithm [29] on the sensitivity of the results and behavior of the cellular automation model. However, the literature does not address the fundamental reasons behind uncertainty generation in PCA. Furthermore, there is a notable lack of exploration concerning their dependence on factors such as temporal and spatial resolution.

In this study, the uncertainty in PCA simulations of microstructural evolution has been investigated. So, the stochastic behavior of microstructure changes in each PCA time step is studied, and discrete probability distribution functions (dPDF) are introduced to analyze the frequency distribution of simulation outcomes. These dPDFs are employed to predict the most probable outcome and the dispersion of simulation results. Also, the effects of modeling parameters, including the size and resolution of the model, as well as the probability distribution of state changes on The precision and reliability of simulation results, are studied. To assess the effectiveness of the suggested dPDFs, different cellular automaton models with various modeling specifications were created, and the results were statistically analyzed.

## 2. Uncertainty in PCA method

In the CA method, the spatial domain is divided into a finite number of cells with specific states $\xi_i^{(k)}$ and neighborhood $\eta_i^{(k)}$ at discrete time step $k$. Also, a state transformation function $f_i^{(k)}$ is defined to compute the new states based on their previous states and neighborhoods. PCA is a type of CA that uses probabilistic rules for state transitions instead of deterministic rules. In this method, the new state of each cell is determined by a Monte Carlo sampling based on the transition probability of the cell and the states of its neighboring cells (it is mathematically expressed in eq. 1). For a cell "$i$" with a transformation probability "$\omega_i$", the cell's state at time step "$k+1$" can be determined by rewriting eq. 1 as follows.

$$\xi_i^{k+1} = \begin{cases} New\ state & if\ \zeta_i^k = 1 \\ \xi_i^k & if\ \zeta_i^k = 0 \end{cases} \tag{2}$$

In this equation, the transformation decision "$\zeta_i^k$" is evaluated using a Monte Carlo sampling, where 1 represents a "success", and 0 means a "failure". As it is clear in eq. 2, there are just two possible outcomes ("success" or "failure") in which $\zeta_i^k = 1$ ("success") occurs with probability $\omega_i$ and $\zeta_i^k = 0$ ("failure") occurs with probability $1 - \omega_i$, where $0 \leq \omega_i \leq 1$. Thus, the Bernoulli probability function $p_i$ over possible outcomes $\zeta_i \in \{1, 0\}$ can be expressed to determine the probability of possible outcome $\zeta_i$ as follows

$$p_i(\zeta_i; \omega_i) = \begin{cases} \omega_i & if\ \zeta_i = 1 \\ 1 - \omega_i & if\ \zeta_i = 0 \end{cases} \quad 0 \leq p_i \leq 1 \tag{3}$$

It may be simply rewritten as follows

$$p_i(\zeta_i; \omega_i) = \zeta_i \omega_i + (1 - \zeta_i)(1 - \omega_i) \quad 0 \leq p_i \leq 1,\ \zeta_i \in \{1, 0\} \tag{4}$$

Fig. 1. schematically depicts the probability distribution of cell $i$.

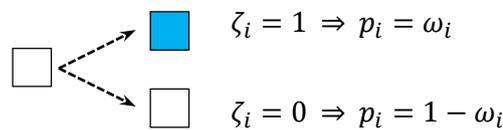

Fig. 1. Probability distribution of cell $i$ within a time step.

In the simulation of microstructural transformations, similar to other PCA applications, many cells are subjected to transformation at each time step. Fig. 2 provides a schematic representation of a microstructure, illustrating two distinct blue and white phases separated by a red boundary. It is assumed that the boundary is moving in the specified directions (local boundary velocities are depicted by red vectors). Consequently, the cells highlighted in red are subjected to transformation.

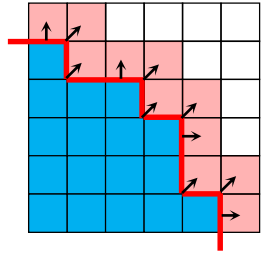

Fig. 2. Schematic representation of a microstructure, illustrating two distinct blue and white phases separated by a red boundary. The cells highlighted in red are subjected to transformation.

In the cellular automaton method, it is crucial to note that the transformation of each cell is determined by the states of the cell and its neighboring cells, independent of the transformations of other cells. In other words, the transformation of each cell is independent of the transformations in other cells. It is assumed that the domain consists of $\mathcal{N}$ cells, and at the $k^{th}$ time step, $N$ cells are subjected to transformation. So, there are $2^N$ possible outcomes or scenarios. Fig. 3 schematically depicts all possible scenarios for the transformation of boundary cells.

| Outcome possibility | Number of Successes (n) | Probability of each scenario |
|---|---|---|
| 1 | 0 | $(1-\omega_1)(1-\omega_2)(1-\omega_3)\ldots(1-\omega_{N-1})(1-\omega_N)$ |
| 2 | 1 | $\omega_1(1-\omega_2)(1-\omega_3)\ldots(1-\omega_{N-1})(1-\omega_N)$ |
| 3 | 1 | $(1-\omega_1)\omega_2(1-\omega_3)\ldots(1-\omega_{N-1})(1-\omega_N)$ |
| ⋮ | ⋮ | ⋮ |
|  | 1 | $(1-\omega_1)(1-\omega_2)(1-\omega_3)\ldots\omega_{N-1}(1-\omega_N)$ |
|  | 1 | $(1-\omega_1)(1-\omega_2)(1-\omega_3)\ldots(1-\omega_{N-1})\omega_N$ |
| ⋮ | ⋮ | ⋮ |
| $2^N-1$ | $N-1$ | $\omega_1\omega_2\omega_3\ldots\omega_{N-1}(1-\omega_N)$ |
| $2^N$ | $N$ | $\omega_1\omega_2\omega_3\ldots\omega_{N-1}\omega_N$ |

Fig. 3. Schematic diagram of all possible scenarios for transformation of boundary cells.

If $n$ represents the number of successes ($\{n \in \mathbb{N}_0 \mid n \leq N\}$), a specific outcome $\mathcal{O}_j$ denoting the $j^{\text{th}}$ possible way to see $n$ successes among $N$ cells can be defined as follows

$$\mathcal{O}_{n,j} = \{\zeta_1, \ldots \zeta_i, \ldots \zeta_N\} \mid \zeta_i \in \{1, 0\} \ \& \ \sum_{i=1}^{N} \zeta_i = n \tag{5}$$

It should be noted that the number of possible scenarios with $n$ successes among $N$ cells is $\mathcal{J} = \binom{N}{n}$ (combinatorial number $N$ in $n$). Hence, the probability of occurring $j^{\text{th}}$ scenario with $n$ successes is

$$\Pr(\mathcal{O}_{n,j}) = \Pr(\cap_{i=1}^{N} \zeta_i) \quad \text{where } \zeta_i \in \mathcal{O}_{n,j}, \ \forall j \in \mathbb{N} \mid j \leq \mathcal{J} \tag{6}$$

As the finite set of events in each scenario is mutually independent, by using the multiplication rule for independent events, yielding

$$\Pr(\mathcal{O}_{n,j}) = \prod_{i=1}^{N} p_i(\zeta_i; \omega_i) = \prod_{i=1}^{N} (\zeta_i \omega_i + (1 - \zeta_i)(1 - \omega_i)) \tag{7}$$

As mentioned earlier, there are $\mathcal{J}$ different scenarios for $n$ successes. Considering all these scenarios, the probability of $n$ successes can be calculated as follows

$$\Phi(n) = \Pr(\cup_{i=1}^{\mathcal{J}} \mathcal{O}_{n,i}) \tag{8}$$

So it can be expanded as below

$$\Phi(n) = \sum_{i=1}^{\mathcal{J}} \Pr(\mathcal{O}_{n,i}) - \sum_{i<j} \Pr(\mathcal{O}_{n,i} \cap \mathcal{O}_{n,j}) + \sum_{i<j<k} \Pr(\mathcal{O}_{n,i} \cap \mathcal{O}_{n,j} \cap \mathcal{O}_{n,k}) + \cdots + (-1)^{\mathcal{J}-1} \sum_{i<\cdots<\mathcal{J}} \Pr(\cap_{i=1}^{\mathcal{J}} \Pr(\mathcal{O}_{n,i})) \tag{9}$$

In this problem, different scenarios are mutually exclusive which means the occurrence of any one of them implies the non-occurrence of the remaining scenarios. As the mutually exclusive scenarios cannot occur at the same time, so

$$\cap_{i \in I} \Pr(\mathcal{O}_{n,i}) = \emptyset \ \ \text{for every } I \subset \{1, \ldots, \mathcal{J}\} \ \& \ |I| > 1 \tag{10}$$

so, eq. 9 can be simplified to

$$\Phi(n) = \sum_{i=1}^{\mathcal{J}} \Pr(\mathcal{O}_{n,i}) = \sum_{i=1}^{\mathcal{J}} \left( \prod_{j=1}^{N} p_i(\zeta_i; \omega_i) \right) \tag{11}$$

This dPDF determines the frequency distribution of successful transformations of boundary cells. So, the most repetitive outcome and its uncertainty at each time step can be assessed using this dPDF. The mean or expected value of the dPDF is calculated as the weighted average of all discrete possible values, and it can be expressed as follows

$$\mu = \sum_{n=0}^{N} n \Phi(n) \tag{12}$$

Also, the squared deviation from the mean (variance) of $\Phi(n)$ can be defined as follows

$$\sigma^2 = \sum_{n=0}^{N} \Phi(n)(n-\mu)^2 \tag{13}$$

This parameter is employed to quantify the dispersion of simulation results, explicitly indicating the level of uncertainty in the obtained outcomes. As a simple case study, it can be assumed that the transformation probabilities of all boundary cells are the same, where $\omega_j = \omega$ for every $j \in \{1, \ldots, N\}$, so eq. 7 can be rewritten as

$$\Pr(\mathcal{O}_{n,j}) = \prod_{\substack{i=1 \\ \zeta_i = 1}}^{N} \omega \prod_{\substack{i=1 \\ \zeta_i = 0}}^{N} (1-\omega) = \omega^n (1-\omega)^{N-n} \tag{14}$$

Consequently, the probability distribution of $n$ successes derived as

$$\Phi(n) = \sum_{i=1}^{\mathcal{J}} \Pr(\mathcal{O}_{n,i}) = \sum_{i=1}^{\mathcal{J}} \omega^n (1-\omega)^{N-n} = \binom{N}{n} \omega^n (1-\omega)^{N-n} \tag{15}$$

It can be observed that the distribution of successful transformations in a given time step follows a binomial distribution. The binomial distribution is a dPDF that models the number of successes in a fixed number of independent Bernoulli trials with the same success probability. This problem is similar to tossing a coin $N$ times, where the probability of getting a "heads" in each toss is $\omega$.

In this case, it can be shown that the mean of the distribution is equal to $N\omega$. Also, the distribution mode, which represents the most probable number of successful transformations among $N$ boundary cells, is an integer number that satisfies the following statement.

$$\text{mode} = \begin{cases} \lfloor (N+1)\omega \rfloor & \text{if } (N+1)\omega \notin \mathbb{N}_0 \\ (N+1)\omega \text{ and } (N+1)\omega - 1 & \text{if } (N+1)\omega \in \mathbb{N} \\ N & \text{if } \omega = 1 \end{cases} \tag{16}$$

Furthermore, the variance of the results distribution can be expressed using the following equation.

$$\sigma^2 = N\omega(1-\omega) \tag{17}$$

## 3. Results & Discussion
### 3.1. Uncertainty Assessment in PCA method

At the end of the previous section, the simplest case of the uncertainty analysis of PCA simulation results is proposed. In this case, it's assumed that within a cellular model consisting of $\mathcal{N}$ cells, there are $N$ boundary cells with a transformation probability of $\omega$. In this condition, the distribution of simulation results can be predicted using the binomial distribution function, as described by Eq. 15. For instance, Fig. 4 illustrates the discrete bell-shaped distributions with different values of $\omega$. The horizontal axis represents the ratio of successful transformations to the total number of cells subject to transformation ($n/N$), while the vertical axis represents the probability of occurrence. Notably, the highest probability of transformation occurs when the ratio $n/N$ is equal or nearly equal to the probability of transformation, as predicted by Eq. 16. Furthermore, it is observed that the highest dispersion of results occurs at $\omega = 0.5$. In contrast, when $\omega = 0$ or $\omega = 1$, no dispersion is observed in the results, representing 0% and 100% transformation success, respectively.

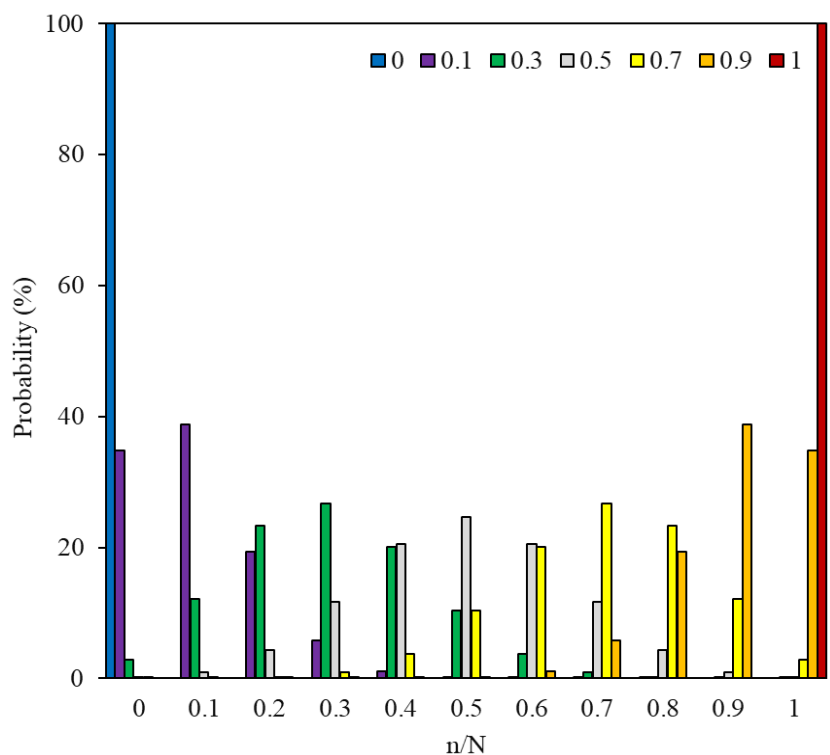

Fig. 4. Binomial distributions, N=10 with different $\omega = $ 0, 0.1, 0.3, 0.5, 0.7, 0.9 and 1.

To evaluate the ability of the binomial distribution function, numerous PCA models were created and employed to achieve the distribution of simulation results. The model consisted of a specific number of boundary cells with a uniform distribution of transformation probability. Also, the probabilistic state change rule described in Eq. 1 was applied to determine cell transformations. To obtain the distribution of the PCA results, the PCA simulation was repeated several times with the same inputs. Fig. 5 illustrates both the histogram of PCA outcomes and the binomial discrete distribution function with 20 boundary cells and $\omega = 0.5$. Notably, the distribution of the PCA model results exhibits a close consistency with the binomial distribution function, highlighting the potency of the binomial model in the prediction of the distribution of the PCA results at each time step. It should be noted that due to the inherent stochastic nature of the algorithm, the repeated PCA simulations may exhibit chaotic behavior, leading to dispersion of the results.

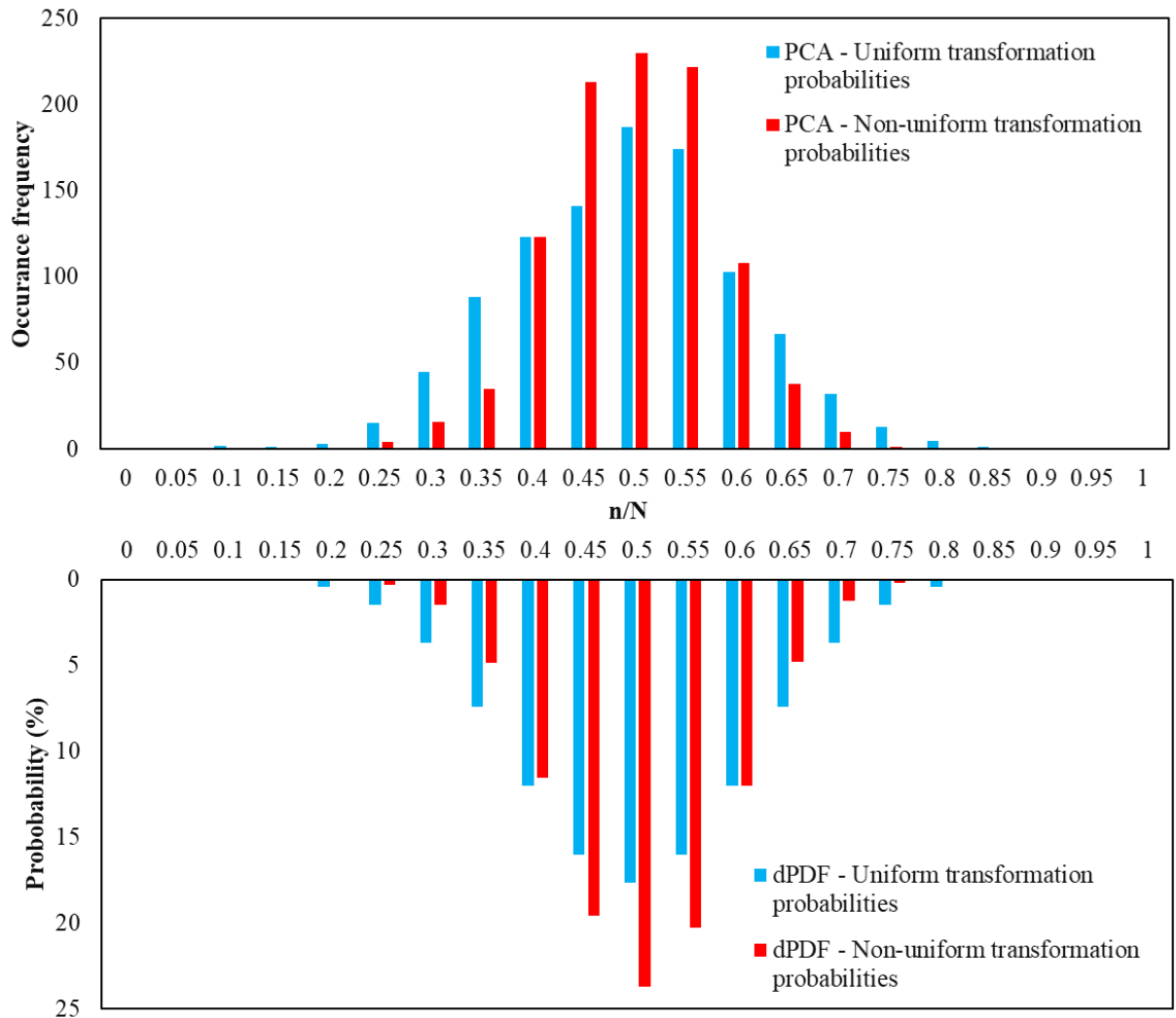

Fig. 5. Comparison of the distribution functions (Bottom histograms, blue histogram: Binomial distribution with $\omega = 0.5$, and red histogram: distribution function with non-uniform cell transformation probabilities $\bar{\omega} = 0.5$) and distribution of PCA results replication (Top histograms, blue histogram: PCA results' distribution with $\omega = 0.5$, and red histogram: PCA results' distribution with non-uniform cell transformation probabilities $\bar{\omega} = 0.5$). Number of replication =1000, N = 20.

The average and the standard deviation of the simulation results can be computed by repeating the PCA simulations with different $\omega$ and obtaining the distribution of the results. Comparisons between the mean value and standard deviation of the simulation results and the binomial distribution are presented in Fig. 6. In Fig. 6a, it can be observed that the mean values of the repeated simulations align well with the mean of the binomial distribution function ($\mu = \omega$). Furthermore, the standard deviation values of the PCA simulations (Fig. 6b) exhibit a notable agreement with the parabolic function as described in Eq. 17.

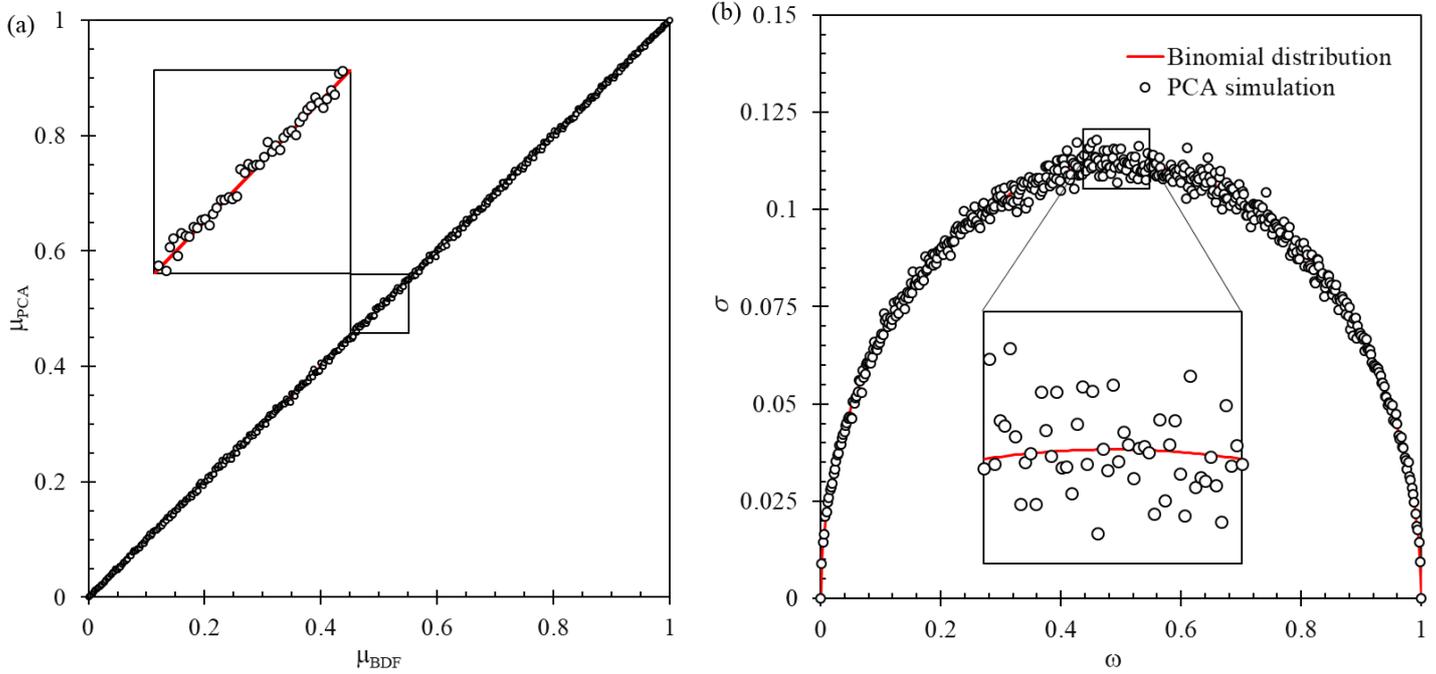

Fig. 6. Comparisons between the mean value and standard deviation of the simulation results and the binomial distribution.

In PCA simulations, the transformation possibilities of boundary cells within the microstructure are not uniform due to the non-uniform distribution of driving pressure and the mobility of boundaries. To assess the accuracy of these simulations, similar to the approach used in the previous section, cellular models with a specific number of boundary cells were employed. However, it is assumed that the transformation probability of the boundary cells is non-uniform. The transformation probabilities of boundary cells were randomly assigned, whereas the average transformation probability is equal to $\bar{\omega}$, the following relationship was satisfied

$$\bar{\omega} = \frac{1}{N}\sum_1^N \omega_i \ \ where \ \omega_i \in \mathbb{R} \mid 0 \leq \omega_i \leq 1 \tag{18}$$

The frequency distribution of the results of repeated PCA simulations is presented in Fig 5. These simulations were conducted with non-uniform cell transformation probabilities with a predetermined $\bar{\omega}$ value of 0.5. Additionally, the probability distribution based on Eq. 11 is compared with the binomial distribution function, and the histogram of PCA simulation results is obtained using a constant transformation probability. The histograms demonstrate a notable agreement among the different distributions, indicating similar patterns in the simulation outcomes. However, it seems that the results obtained under non-uniform transformation probabilities exhibit a narrower distribution than those under uniform conditions. In other words, the modeling and simulation outcomes under non-uniform transformation probabilities show smaller dispersion compared to the uniform condition. Further statistical analysis can provide additional insights into the degree of agreement among the distributions.

Fig. 7 illustrates the mean and standard deviation of the distribution of PCA simulation results, and the distribution function represented in eq. 11 with different $\bar{\omega}$. In this figure, the circles represent the results of eq. 11, and the plus symbols show the PCA simulation results. This figure reveals the ability of eq. 11 to predict the distribution of PCA simulation results, as evidenced by the consistent alignment of the statistical parameters of the simulation results and the distribution function predictions. Fig. 7a illustrates the relationship between the mean of the distribution function and $\bar{\omega}$. So, it may be concluded that the mean of the distribution function equals $N\bar{\omega}$. In other words, the mean of the distribution function in Eq. 12 is equivalent to the sum of the transformation probabilities of boundary cells.

$$\mu = \sum_{n=0}^{N} n\Phi(n) = \sum_{i=1}^{N} \omega_i \tag{19}$$

The standard deviation of the results from repeated PCA simulations and the distribution function presented in Eq. 13 are displayed in Fig. 7b. To calculate the distribution function using Eq. 13, all possible events should be considered, which can be computationally intensive. As discussed in section 2, the number of possible events in each time step follows an exponential relation (number of events equal $2^N$) with the number of boundary cells ($N$) and leads to a substantial increase in computational demands with larger $N$. Consequently, using eq. 13 to calculate the standard deviation becomes less feasible in cases with a large number of boundary cells. Therefore, an alternative solution is needed to resolve this limitation.

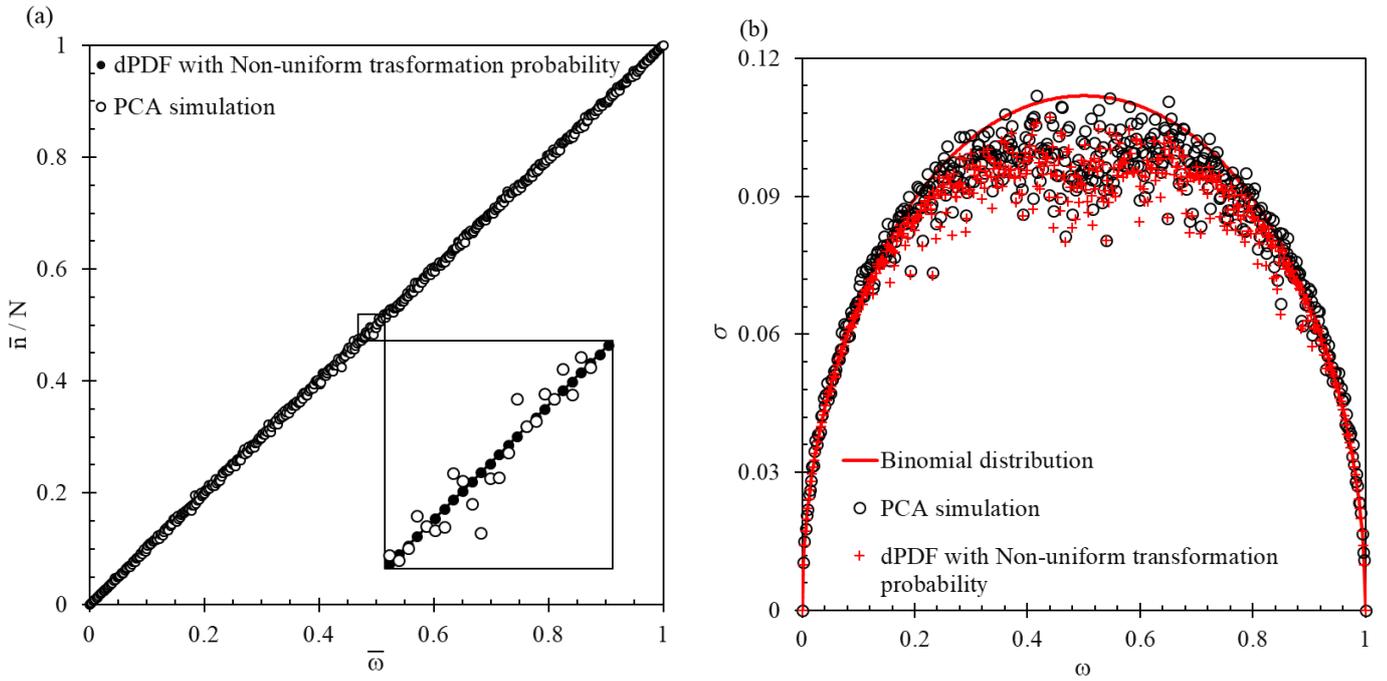

Fig. 7. Comparison of (a) the mean and (b) standard deviation of the distribution of PCA simulation results, and the dPDF represented in eq. 11 in different $\bar{\omega}$.

In the previous sections, it was observed that when the transformation probability of boundary cells is uniform, the binomial distribution function can effectively predict the distribution of the results. However, to provide an alternative solution, $\bar{\omega}$ (the mean of the transformation probability of boundary cells) was inserted into the binomial distribution function (equations 15 & 17). The solid line in Fig. 7 demonstrates the performance of this modified distribution. As depicted in this figure, the standard deviation obtained from the binomial function provides a reasonable approximation of the standard deviation derived from PCA simulations and the distribution function of Eq. 13. Notably, the predicted standard deviation using the binomial distribution function is higher than standard deviations of almost all PCA simulations and all results of the distribution function of Eq. 13. In other words, the upper limit of the standard deviation and the maximum uncertainty in the results can be estimated using the binomial distribution function, therefore

$$\sigma^2 \leq N\bar{\omega}(1-\bar{\omega}) \tag{20}$$

The above equation is one of the key findings of this research, offering a simple and computationally efficient method to forecast the uncertainty of PCA simulation results at each time step.

## 3.2. PCA Model Parameters and Uncertainty of Results

In this section, Eq. 20 is employed to investigate the effects of various modeling parameters, including the number of spatial dimensions, cellular resolution, and model size, as well as the average transformation probability of boundaries on uncertainty in simulation results. In this section, to assess the dispersion of results compared to the best estimate of PCA simulation outcomes, the coefficient of variation is used. This dimensionless factor is obtained by calculating the ratio of the standard deviation to the mean value of the results ($C_v = \sigma / \mu$) and provides insight into the precision and reliability of the simulation results. This factor quantifies the relative variability in the simulation results and allows meaningful comparisons of different simulations. Using eq. 18-20, the coefficient of variation "$C_v$" can be derived as follows

$$C_v \leq \frac{(N\bar{\omega}(1-\bar{\omega}))^{\frac{1}{2}}}{N\bar{\omega}} = \left(\frac{1-\bar{\omega}}{N\bar{\omega}}\right)^{\frac{1}{2}} \qquad (21)$$

The right-hand term of this equation represents the upper bound of $C_v$. The dependency of the upper limit of $C_v$ on $\bar{\omega}$ and the number of boundary cells is illustrated in Fig. 8. It is evident that when $\bar{\omega}$ equals 1, the PCA simulation results exhibit no dispersion. It indicates complete success in transformations of boundary cells without any errors or uncertainties. However, as $\bar{\omega}$ decreases, an increase is observed in the relative dispersion of the simulation results. When $\bar{\omega}$ approaches zero, the intensity of the upper bound of $C_v$ increases significantly. At low values of $\bar{\omega}$, where the probability of transformation in boundary cells is very low, even the transformation of a single cell causes significant results' scattering compared to the mean value of the results. It's important to note that according to Fig. 6 and 7, the highest uncertainty occurs at $\bar{\omega} = 0.5$. Moreover, the figure reveals that increasing the number of boundary cells reduces the coefficient of variation and the dispersion of the results. In other words, by increasing the number of boundary cells, the uncertainty of the results is reduced, leading to enhanced repeatability of the PCA simulation results. These findings may be used to determine and optimize the reliability and precision of the simulations.

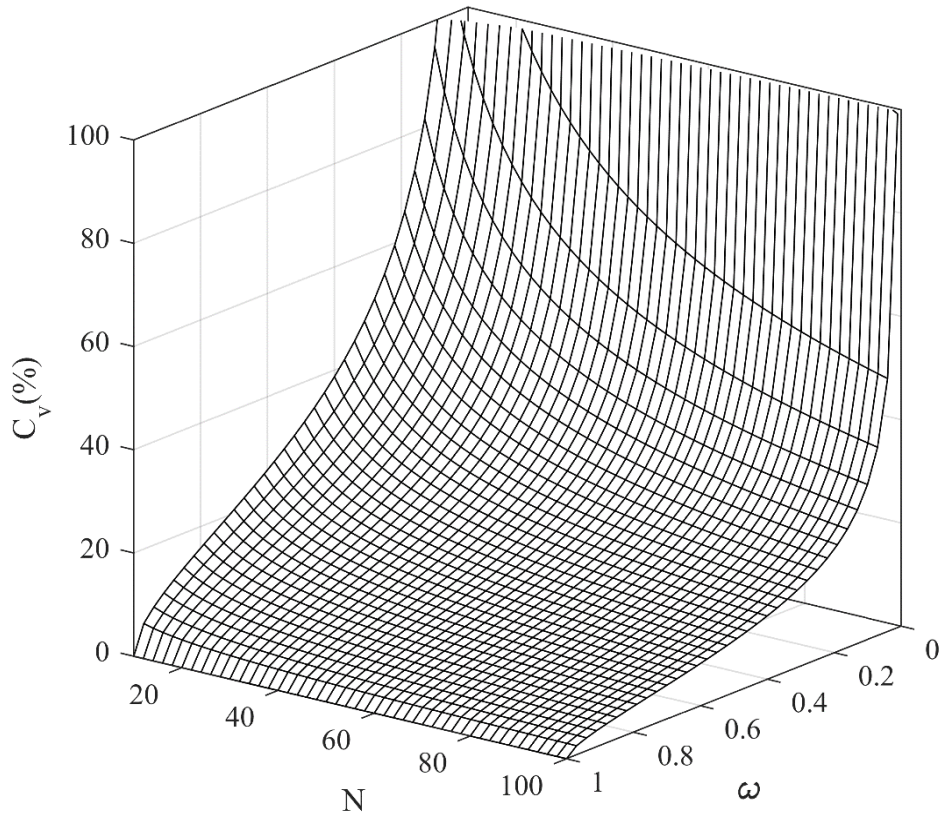

Fig. 8. The dependency of the upper limit of coefficient of variation "$C_v$" on the mean of the transformation probability of boundary cells "$\bar{\omega}$" and the number of boundary cells "$N$".

A comparison of the upper limit of Cv calculated using eq. 21 with PCA simulation results with different numbers of boundary cells is presented in Fig. 9. Remarkably, the figure illustrates that eq. 21 effectively predicts the upper limit of Cv and demonstrates the potential of this equation in exploring the effects of various modeling parameters on the uncertainty of PCA simulations. Fig. 9b illustrates the case with $\bar{\omega} = 0.5$. A noteworthy observation is the drastic decrease in the change rate of Cv with increasing the number of boundary cells. In other words, PCA simulations with a large number of boundary cells exhibit less sensitivity to further increases in $N$. This observation is significant as it highlights the influence of the number of boundary cells on the stability and reliability of simulation results.

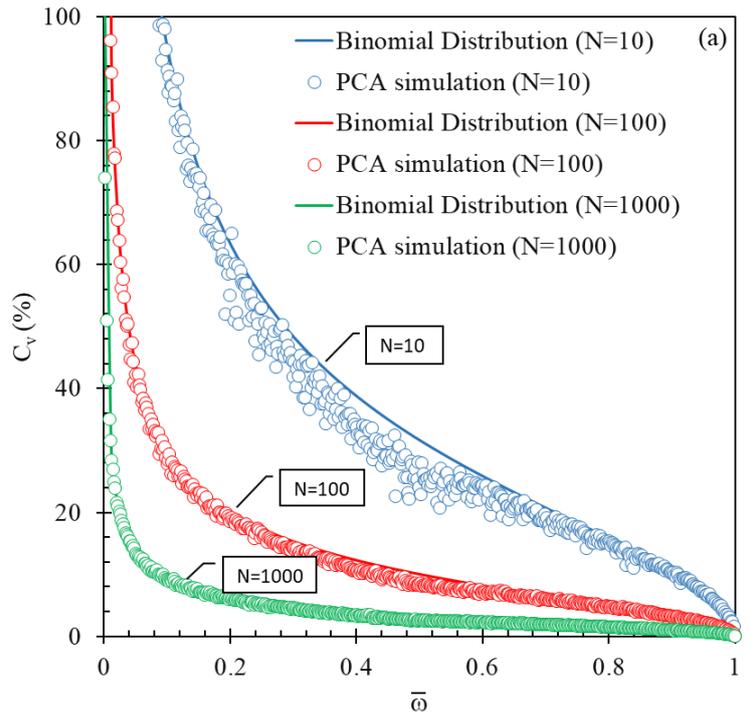

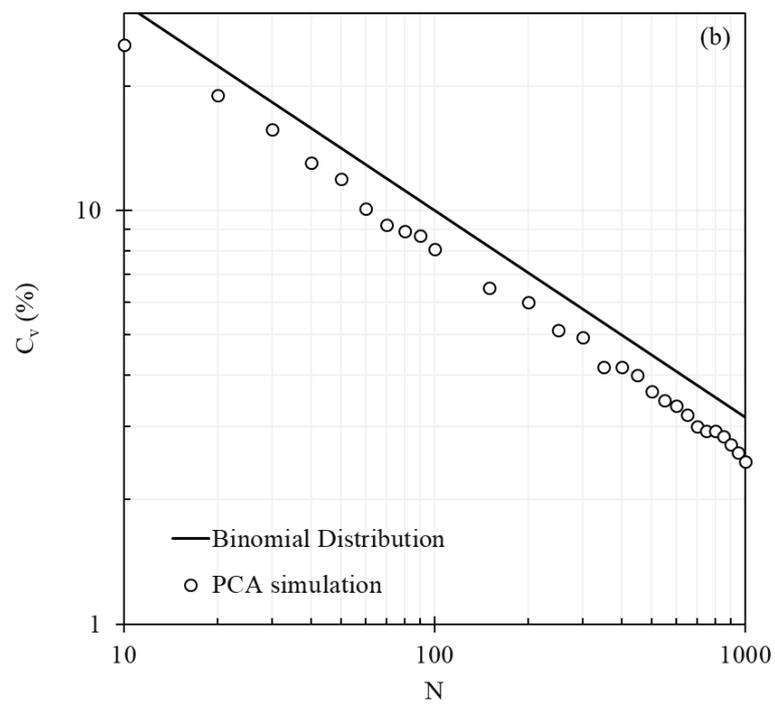

Fig. 9. Comparison of the upper limit of Cv calculated using eq. 21 with PCA simulation results with different numbers of boundary cells.

In cellular automata simulations of microstructural transformations, the number of boundary cells depends on the modeling parameters such as model size and cell resolution (number of cells per unit length). Generally, for a given microstructure morphology, by increasing the model size and cell resolution, both the area of grain boundaries and the number of boundary cells increase. For instance, the cellular microstructures with the same average grain size but different model sizes and cellular resolutions in two- and three-dimensional are presented in Fig. 10 and 11, respectively. Notably, increasing the model size and cell resolution leads to an increase in the number of boundary cells. In the subsequent analysis, the effects of these parameters on the dispersion of PCA simulation results are analyzed.

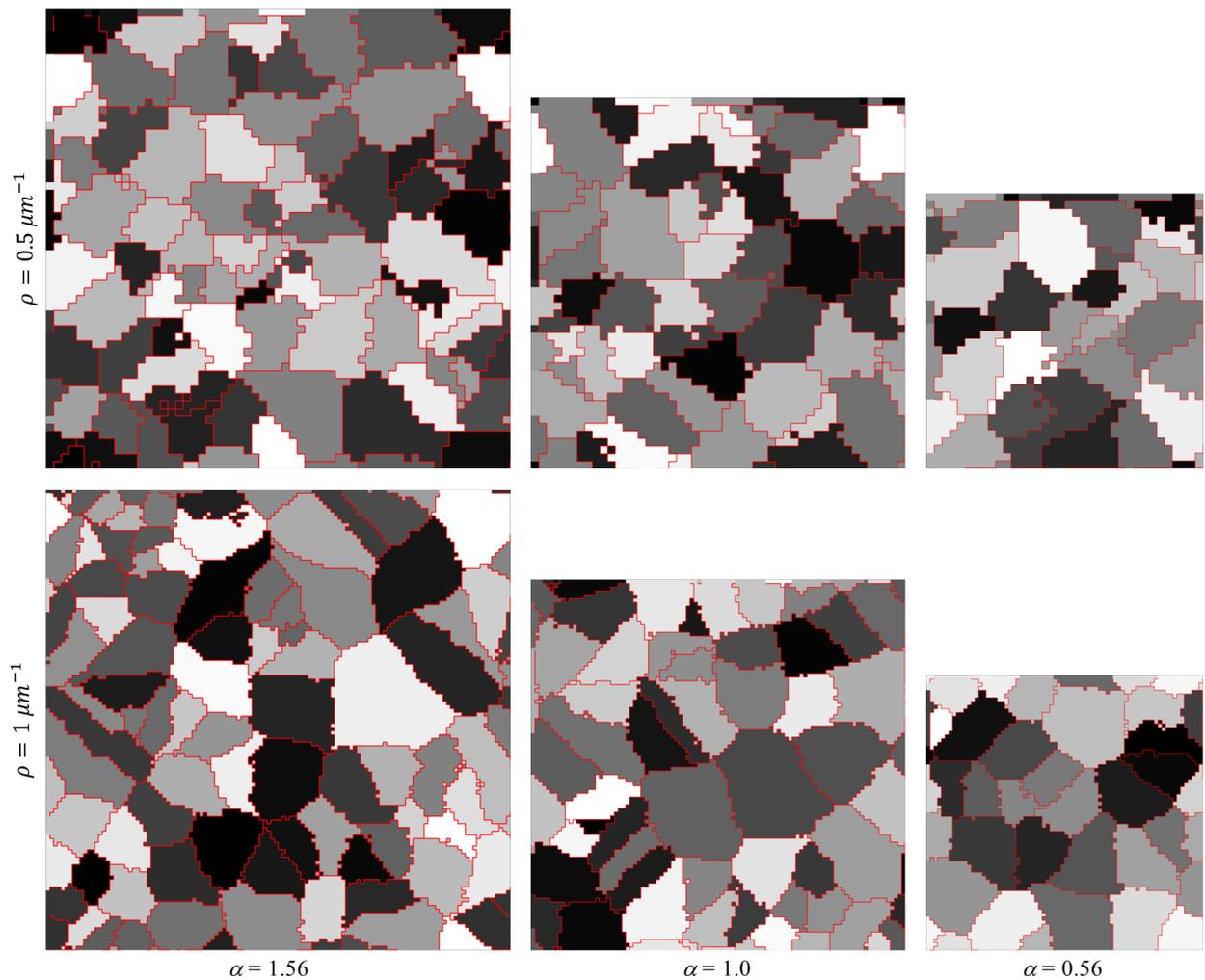

Fig. 10. Two-dimensional cellular microstructures with the same average grain size but different model sizes and cellular resolutions.

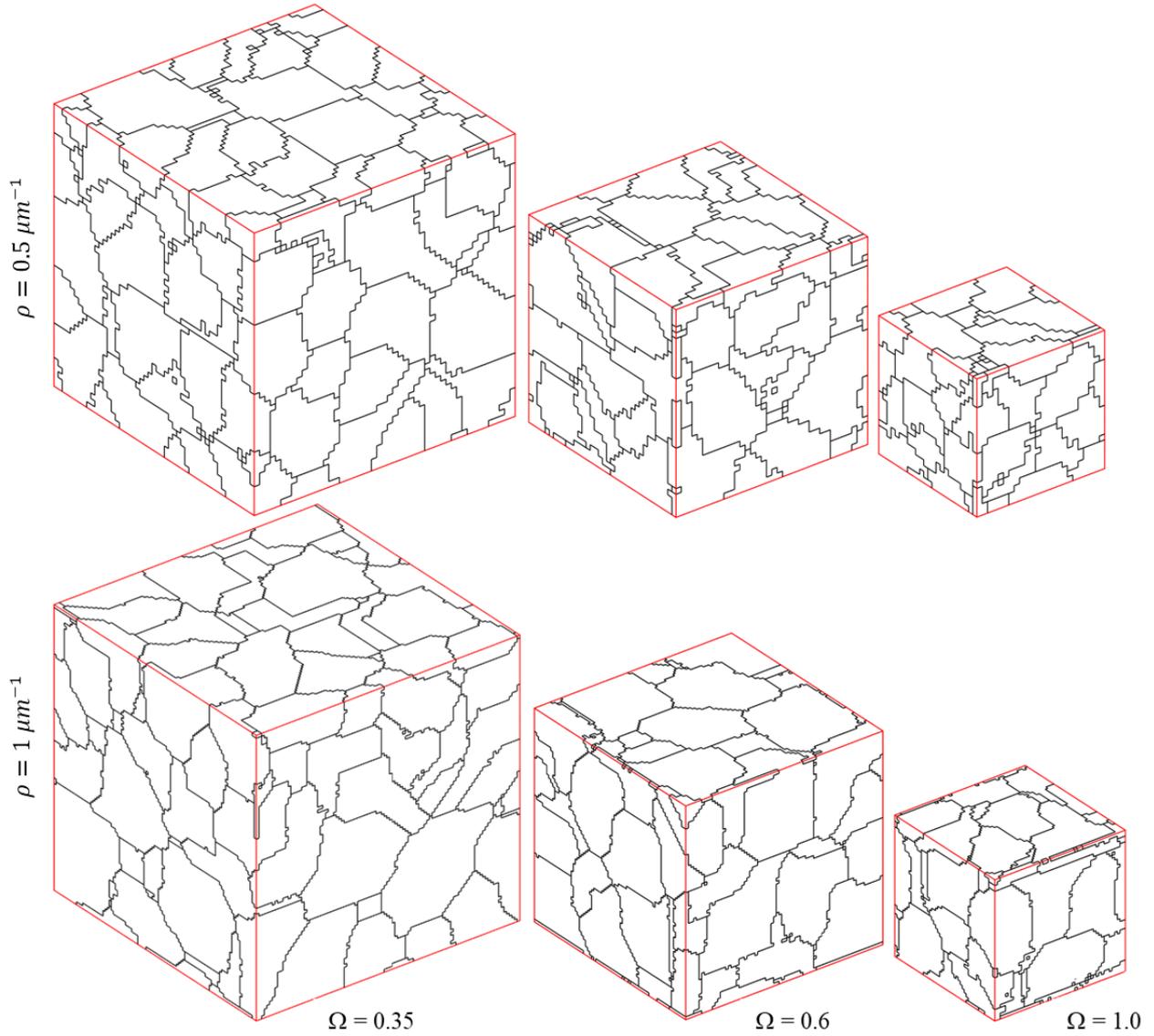

Fig. 11. Three-dimensional cellular microstructures with the same average grain size but different model sizes and cellular resolutions.

In 2D cellular automaton modeling, adjusting the model's size with a specific ratio statistically leads to a proportional change in the length of the grain boundaries. In other words, if the boundary length of a microstructure with a specific model's size is denoted as $L_0$, by changing the model's size with a ratio of "$\alpha$", the boundary length of the new model can be estimated as $\alpha L_0$. Additionally, the number of boundary cells can be estimated as $\rho L$ (where $L$ represents the length of the grain boundaries and $\rho$ is cell resolution). Thus, the eq. 21 can be rewritten as follows.

$$C_v \leq \left(\frac{1-\bar{\omega}}{\alpha \rho L_0 \bar{\omega}}\right)^{\frac{1}{2}} \qquad (22)$$

Fig. 12a presents the comparison between the predicted upper limit of $C_v$ using eq. 22 and 2D PCA simulation results in various cell resolutions and model sizes. Throughout this analysis, it is assumed that $\bar{\omega}$ remains constant regardless of changes in model size and cell resolution. Under this assumption, $C_v$ can be derived as the following relationship. As demonstrated in Fig. 12a, increasing the cell resolution and employing larger models result in a reduction of uncertainty in the simulation results. The findings showcased in this figure provide valuable insights into the influence of cell resolution and model size on the reliability and stability of the PCA simulations. By optimizing these parameters, the precision and reproducibility of the microstructure modeling process can be enhanced. The following equation shows the effects of model size and resolution on the variation of $C_v$ compared to a specific model with resolution of $\rho_0$ and uncertainty of $C_{v0}$

$$C_v = \left(\frac{\rho_0}{\alpha\rho}\right)^{\frac{1}{2}} C_{v0} \tag{23}$$

For instance, $C_v$ can be halved by doubling each dimension of the model (resulting in four times the area of the 2D model) or quadrupling the cell resolution. In 3D PCA modeling, a similar analysis to that presented in the previous section allows us to explore the relationship between PCA simulation uncertainty and model size, as well as cell resolution. In 3D modeling, the area of the grain boundaries also changes proportionally by changing the size of the model. Furthermore, as the grain boundaries in 3D modeling are spatial surfaces with a specific area, the number of boundary cells is related to the square of the cell resolution. Consequently, the upper bound of $C_v$ follows the relationship presented below

$$\frac{\sigma}{\bar{n}} \leq \left(\frac{1-\bar{\omega}}{N\bar{\omega}}\right)^{\frac{1}{2}} = \left(\frac{1-\bar{\omega}}{\rho^2 \Omega S_0 \bar{\omega}}\right)^{\frac{1}{2}} \tag{24}$$

where $\Omega$ represents the volume ratio of the adjusted model compared to the original model. Fig. 12b shows the comparison between the predicted upper limit of $C_v$ using eq. 24 and 3D PCA simulation results in various cell resolutions and model sizes. As a result, the value of $C_v$ can be expressed as a function of cell resolution and model size as follows

$$C_v = \frac{\rho_0}{\rho\sqrt{\Omega}} C_{v0} \tag{25}$$

For instance, doubling the cell resolution leads to halving the value of $C_v$. Similarly, multiplying each dimension of the model by $4^{1/3}$ yields a similar outcome.

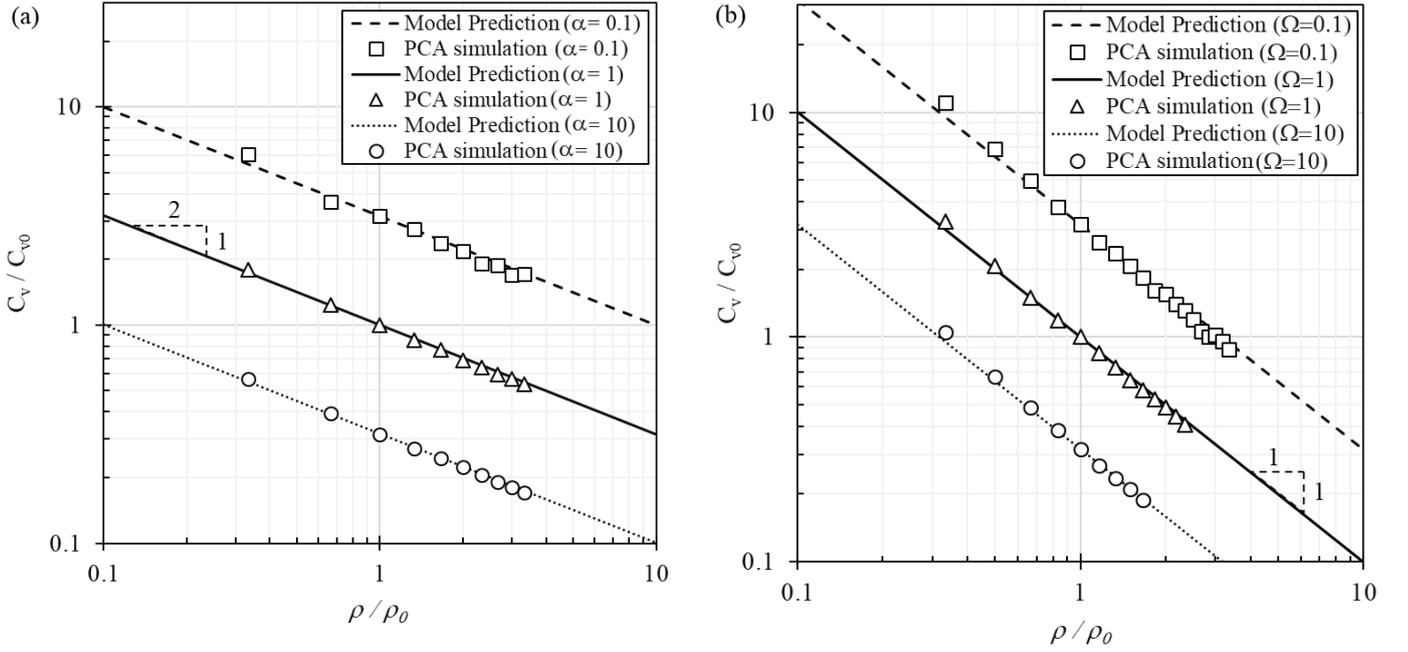

Fig. 12. Comparison between the predicted $C_v$ using eqs. 22 and 24 and PCA simulation results in various cell resolutions and model sizes in (a) 2D and (b) 3D PCA modeling.

## 4. Conclusions

In this study, the uncertainty analysis and verification of Probabilistic Cellular Automata (PCA) simulations in microstructural evolution is investigated. The findings offer practical tools and equations to assess the reliability and precision of the microstructure modeling. The key findings of the investigation can be summarized as follows:

- A discrete probability distribution function is introduced to predict the frequency distribution of simulation results at each PCA time step, offering a key tool for evaluating outcomes and measuring result dispersion. This dPDF can be simplified into binomial distribution in PCA modeling with uniform transformation probabilities for boundary cells.

- The comparison between the statistical analyses of PCA simulation results and the binomial distribution function shows that the variance of simulation outcomes is less than the variance of the binomial distribution function. As a result, The equation $\sigma^2 \leq N\bar{\omega}(1-\bar{\omega})$ offers a straightforward and computationally efficient method to estimate the maximum uncertainty in PCA simulation results at each time step.

- The dPDF predictions and PCA simulation results showed that the maximum dispersion of results was observed in PCA models with an average transformation probability of 0.5.

- the coefficient of variation was introduced as a measure of precision and reliability, its dependence on the average transformation probability and the number of boundary cells. This observation highlights the influence of the number of boundary cells on the stability and reliability of simulation results, indicating that an increase in boundary cells led to more precise outcomes.
- As cellular resolution and model size increase, representing the microstructure with more cells results in a reduction of uncertainty and dispersion in simulation results.
- In 2D PCA models, the relative uncertainty of simulation results was found to be inversely proportional to the square root of cellular resolution and model size. Also, in 3D models, the relative uncertainty of simulation results is inversely proportional to the cellular resolution and square root of model size.

**Data Availability**

The data that support the findings of this study are publicly available at "Seyed Salehi, Majid (2024), Uncertainty Assessment of Probabilistic Cellular Automata in Microstructure Evolution - Data and Resources, Mendeley Data, V1, doi: 10.17632/3p4pfsk3zd.1".